\documentclass[aps,twocolumn,prd,showpacs,nofootinbib]{revtex4}
\usepackage{amsmath}
\usepackage{graphicx}
\usepackage{dcolumn}
\usepackage{bm}
\usepackage{amssymb}
\usepackage{latexsym}

\def\be{\begin{equation}}
\def\ee{\end{equation}}
\def\ba{\begin{eqnarray}}
\def\ea{\end{eqnarray}}

\begin{document}

\title{A Galileon Design of Slow Expansion  }

\author{Zhi-Guo Liu }
\author{Jun Zhang }
\author{Yun-Song Piao}

\affiliation{College of Physical Sciences, Graduate University of
Chinese Academy of Sciences, Beijing 100049, China}

\begin{abstract}

We show a model of the slow expansion, in which the scale
invariant spectrum of curvature perturbation is adiabatically
induced by its increasing mode, by applying a generalized Galileon
field. In this model, initially $\epsilon\ll -1$, which then is
rapidly increasing, during this period the universe is slowly
expanding. There is not the ghost instability, the perturbation
theory is healthy. When $\epsilon \sim -1$, the slow expansion
phase ends, and the available energy of field can be released and
the universe reheats. This scenario might be a viable design of
the early universe.

\end{abstract}

\maketitle

\section{Introduction}

The observations imply that the primordial curvature perturbation
is scale invariant. Thus how generating it has been still a
significant issue, especially for single field. The curvature
perturbation on large scale consists of a constant mode and a mode
dependent of time \cite{Mukhanov}. When one of which is dominated
and scale invariant, the spectrum of curvature perturbation will
be scale invariant. When the scale factor is rapidly changed while
$\epsilon$ is nearly constant, the constant mode is responsible
for that of inflation \cite{G},\cite{LAS},\cite{S1},\cite{MC},
while the increasing mode is for the contraction with matter
\cite{Wands99},\cite{FB},\cite{SS}, both are dual \cite{Wands99}.

In principle, the increasing mode of metric perturbation, which is
scale invariant for $\epsilon\gg 1$ \cite{GKST} or $\epsilon \ll
-1$ \cite{PZhou}, might dominate the curvature perturbation. The
constant mode of metric perturbation is same with the constant
mode of curvature perturbation. The duality of scale invariant
spectrum of metric perturbation has been discussed in
\cite{LST},\cite{Piao0404},\cite{Lid}. The evolution with
$\epsilon\gg 1$ is the slowly contracting, which is that of
ekpyrotic universe \cite{KOS}. While $\epsilon \ll -1$ gives the
slow expansion \cite{PZhou}, which has been applied for island
universe \cite{island}. In certain sense, in Ref.\cite{PZhou} it
was for the first time observed that the slow expansion might
adiabatically generate the scale invariant spectrum of curvature
perturbation, see \cite{Piao0706} for that induced by the entropy
perturbation.

When the available energy of field is released, the slow expansion
phase ends and the universe reheats. Thus the slow expansion might
be a viable scenario of the early universe. In principle, when
$\epsilon$ is constant, whether the increasing mode of the metric
perturbation can be inherited by the curvature perturbation
depends of the physics around the exiting \cite{DH}. However, when
$\epsilon$ is rapidly changed, the thing is altered, see
\cite{KS1} for that of the slow contraction. During the slow
expansion, the scale invariant curvature perturbation can be
naturally induced by its increasing mode \cite{Piao1012}, or its
constant mode \cite{KM},\cite{JK}.


The perturbation mode can leave the Hubble horizon during the slow
expansion requires $\epsilon<0$ \cite{PZhou},\cite{Piao1012}, or a
period after it is required to extend the perturbation mode out of
the Hubble horizon \cite{KM}. Thus in
\cite{PZhou},\cite{Piao1012}, the phantom was applied for a
phenomenological studying. However, there is a ghost instability.
Thus it was argued that the evolution of $\epsilon<0$ emerges only
for a period, the phantom field might be only a simulation of a
full theory without the ghost below certain physical cutoff
\cite{CJM}.

Recently, the cosmological application of Galileon,
\cite{NRT},\cite{DEV}, or its nontrivial generalization
\cite{Vikman},\cite{KYY},\cite{DGS}, has acquired increasing
attentions
\cite{G1},\cite{Tsujikawa},\cite{Sami},\cite{Seery},\cite{Koyama}.
It has been found for generalized Galileon that $\epsilon<0$ can
be implemented stably, there is not the ghost instability. We, in
this paper, will show a model of the slow expansion given in
\cite{Piao1012}, by applying a generalized Galileon field. In this
model, the perturbation theory is healthy, the scale invariant
curvature perturbation is given by itself increasing mode, which
can be consistent with the observations. As will be argued, this
in certain sense validates the argument and calculations in
\cite{PZhou},\cite{Piao1012}

The models of early universe, builded by applying generalized
Galileon, have been studied. In Ref.\cite{KYY}, the inflation
model is implemented by using generalized Galileon field. However,
here what we discuss is an alternative to inflation. There is a
slightly similar scenario in \cite{CNT}. However, in \cite{CNT},
the adiabatic perturbation is not scale invariant, the scale
invariant curvature perturbation is obtained by the conversion of
the perturbations of other light scalar fields. Here, we will see
how the adiabatic perturbation is naturally scale invariant.


\section{As A General Result}
We begin with a brief review on the slowly evolving model in
\cite{Piao1012}. The quadratic action of the curvature
perturbation $\cal R$ is \be S_2\sim \int d\eta d^3x {a^2Q\over
c_s^2}\left({{\cal R}^\prime}^2-{c_s^2}(\partial {\cal
R})^2\right), \ee which is actually general for single field, like
$P(X,\varphi)$ \cite{GM}, generalized Galileon
\cite{Vikman},\cite{KYY},\cite{KYY2}, and the modified gravity
\cite{CHC},\cite{FT}. $Q$ and $c_s^2$ are generally different for
different models. However, $Q>0$ and $c_s^2>0$ should be satisfied
to avoid the ghost and gradient instabilities.

The equation of $\cal R$ is \cite{Muk},\cite{KS}
\be u_k^{\prime\prime} +\left(c^2_s k^2-{z^{\prime\prime}\over
z}\right) u_k = 0, \label{uk}\ee  after defining $u_k \equiv
z{\cal R}_k$, where $'$ is the derivative for $\eta$, $z\equiv
{a\sqrt{2M_P^2 Q}/ c_s}$
. We here only care
the case with constant $c_s^2$. 
When $k^2\ll z^{\prime\prime}/z$, the solution of $\cal R$ given
by Eq.(\ref{uk}) is
\ba {\cal R} & \sim & C\,\,\,\,\, is\,\,\,{{\rm constant}}\,\,\,{
{\rm mode}}\label{C}\\ &or &\, D\int {d\eta\over z^2}\,\,\,\,\,
is\,\,\,{{\rm changed}}\,\,\,{ {\rm mode}} , \label{D}\ea where
$D$ mode is increasing or decaying dependent of different
evolutions.

The scale invariance of $\cal R$ requires ${z^{\prime\prime}\over
z}\sim {2\over (\eta_*-\eta)^2}$,
which implies \ba z\sim {a\sqrt{Q}\over c_s} &\sim & {1\over
\eta_*-\eta}\,\,\, {for}\,\,\, {{\rm constant}}\,\,\,{ {\rm mode}}
\label{z2}\\
&or & (\eta_*-\eta)^2 \,\,\, {for}\,\,\,{ {\rm
increasing}}\,\,\,{{\rm mode}} \label{z1}\ea has to be satisfied,
where initially $\eta\ll -1$. In certain sense, both evolutions
are dual \cite{Wands99}. The results will be different if $c_s^2$
is changed, however, which we will not involve here. In principle,
both $a$ and $Q$ can be changed, and together contribute the
change of $z$. However, only one among them is changed while
another is hardly changed might be interesting, e.g. the
inflation, given by (5), or the contraction dominated by the
matter, given by (\ref{z1}), in which $a$ is rapidly changed while
$Q$ is hardly changed.

However, the case can also be inverse.
When $Q$ is rapidly changed while $a$ is hardly changed, the scale
invariant spectrum of curvature perturbation can also be induced
by either its constant mode \cite{KS1},\cite{KM},\cite{JK}, given
by (5), or its increasing mode \cite{Piao1012}, given by
(\ref{z1}). Though both cases give the scale invariant spectrum,
both pictures are distinct. In general, for the picture in
\cite{Piao1012}, initially $|\epsilon|\gg 1$, which then is
rapidly decreasing, the slow evolution of the scale factor ends
when $|\epsilon|\sim 1$. While for that in
\cite{KS1},\cite{KM},\cite{JK}, initially $|\epsilon|\lesssim 1$,
which then is rapidly increasing. In addition, for
\cite{KS1},\cite{KM},\cite{JK}, during the slow evolution, the
perturbation mode is actually still inside the Hubble horizon.
Thus a period after it is required to extend the perturbation mode
out of the Hubble horizon, while in \cite{Piao1012}, the
perturbation mode can naturally leave the Hubble horizon during
the slow evolution. There is also not the problem pointed in
\cite{LMV}.

Here, we will discuss that in \cite{Piao1012}. We have generally
$Q= \epsilon$ for single field action $P(X,\varphi)$ \cite{GM}.
While the case is slightly complex for generalized Galileon
\cite{Vikman},\cite{KYY}. However, as will be showed in following
section, we actually have $Q\sim |\epsilon|$.

Thus $Q= |\epsilon|$ will be set for general discussions in the
following. In principle, $|\epsilon|$ is dependent of $a$.
However, it can be observed that $a$ is nearly constant for
$|\epsilon|\gg 1$. Thus for (\ref{z1}), we have \be Q= {|\epsilon
|}\sim
\Lambda^4_*(t_*-t)^4, \label{e2}\ee 
since $\eta\sim t$, where
$\Lambda_*$ is $1/t_*$ dimension. 
The Hubble parameter is given by \be H \sim {1\over
\Lambda_*^4(t_*-t)^5}. \label{h}\ee Thus $a$ is given by \be
|\ln\left({a\over a_*}\right)| \sim  {1\over
\Lambda^4_*(t_*-t)^4}\sim {1\over |\epsilon|}. \label{aa}\ee When
initially $\Lambda_*(t_*-t)\gg 1$, i.e.$|\epsilon|\gg 1$, the
evolution corresponds to the slow expansion for $\epsilon\ll -1$,
or the slowly contraction for $\epsilon\gg 1$, since $a/a_*\simeq
1$. The slow evolution ends when $\Lambda_*(t_*-t) \simeq 1$, at
which $|\epsilon|\sim 1$.

\begin{figure}[t]
\includegraphics[width=7cm]{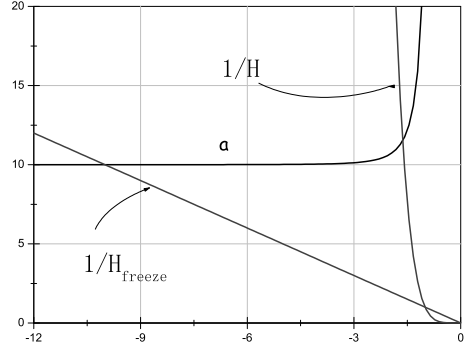}
\caption{The evolutions of $a$, the Hubble horizon and the $\cal
R$ horizon during the slow expansion given by Eq.(\ref{e2}).
$a_*=10$ is set. During this phase, due to the rapidly change of
$H$ and $H_{freeze}$, the perturbation mode initially inside both
horizons, i.e. $\lambda\sim a\ll 1/H_{freeze}\ll 1/H$ will
naturally leave the $\cal R$ horizon, i.e.$\lambda\sim a>
1/H_{freeze}$, and then the Hubble horizon, i.e.$\lambda\sim a>
1/H$. }
\end{figure}

When $k^2\simeq z^{\prime\prime}/z$, the perturbation mode is
leaving the horizon, and hereafter it freezes out. This horizon
might be called as the $\cal R$ horizon \be 1/{\cal
H}_{freeze}=\sqrt{\left|{z\over z^{\prime\prime}}\right|}\simeq
\eta_*-\eta. \ee Thus the physical $\cal R$ horizon is $a/{\cal
H}_{freeze}\simeq t_*-t$, While the Hubble horizon is $1/H$ given
by Eq.(\ref{h}). Here, the evolutions of the $\cal R$ horizon and
the Hubble horizon are different. While when $a$ is rapidly
changed and $|\epsilon|$ is unchanged, e.g.inflation, both
evolutions are almostly same. The reason is that for inflation,
$z^{\prime\prime}/z\sim a^{\prime\prime}/a$, thus \be 1/{\cal
H}_{freeze}\simeq \sqrt{\left|{z\over
z^{\prime\prime}}\right|}\simeq \sqrt{\left|{a\over
a^{\prime\prime}}\right|}\sim 1/{\cal H}, \ee while here $a$ is
constant and $|\epsilon |$ is rapidly changed, we have not
$z^{\prime\prime}/z\sim a^{\prime\prime}/a$.

When $k^2\gg z^{\prime\prime}/z$, i.e. the perturbation is deep
inside the $\cal R$ horizon, $u_k$ oscillates with a constant
amplitude. The quantization of $u_k$ is well defined for $Q\sim
|\epsilon|>0$, which gives its initial value.
The evolutions of $a$, $1/H$ and $a/{\cal H}_{freeze}$ are plotted
in Fig.1 for the slow expansion. It can be found that the
perturbation mode firstly leaves the $\cal R$ horizon, after which
it is freezed out, but it is still inside the Hubble horizon.
However, since the Hubble horizon is decreasing, after a while the
perturbation mode will be inevitably extended outside it, and
become the primordial perturbation on super Hubble scale.

When $k^2\ll z^{\prime\prime}/z$, the amplitude of perturbation
spectrum is $ {\cal P}^{1/2}_{\cal R} \simeq
\sqrt{k^3}\left|{u_k\over z}\right|$. Thus  \be {\cal P}_{\cal R}
\simeq
{{|\epsilon|}\over c_s
M_P^2}{H^2}, \label{P2}\ee where $Q\sim |\epsilon|$ is applied.
The perturbation is given by the increasing mode (\ref{D}),
because $a$ is hardly changed and $|\epsilon|$ is decreasing. When
$|\epsilon|\sim 1$, the change of $a$ begins to become not
negligible. Though $|\epsilon|$ is still decreasing, $a$ is
increased exponentially. Thus this mode will become the decaying
mode at certain time $t_{f}\sim {\cal O}(t_*)$ shortly after
$|\epsilon|\sim 1$.
In principle, the spectrum of $\cal R$ should be calculated around
$t_{f}$.
Thus \be {\cal P}^{1/2}_{\cal R} \sim \sqrt{1\over c_s
M_P^2}H_{f}. \label{P3}\ee
The universe reheats around or after $t_{f}$, and hereafter the
perturbation is dominated by its constant mode, until it enters
into the Hubble horizon during the radiation or matter domination.
$ |\epsilon_{f}| \sim 1$ brings $ \Lambda_*^4(t_*-t_{f})^4 \sim
1$. Thus Eq.(\ref{P3}) becomes \be {\cal P}^{1/2}_{\cal R} \sim
{\Lambda_*\over M_P\sqrt{c_s}}, \label{P1}\ee which is general
result of the slow evolution in \cite{Piao1012}, i.e. the
evolution of $|\epsilon|$ follows Eq.(\ref{e2}) and $c_s^2$ is
constant.


\section{A Galileon Design of Slow Expansion}

Here, we will detailed show a model of the slow expansion given in
\cite{PZhou},\cite{Piao1012}. While the scenario of the slow
contraction given in \cite{Piao1012} is slightly alike with that
in \cite{KS1}, which might be studied in detail elsewhere.

\subsection{The background}

We consider a generalized Galileon as \be {\cal L}\sim
-\,e^{4\varphi/{\cal M}}\,X+{1\over {\cal M}^8}X^3-{1\over {\cal
M}^7}X^2\Box\varphi , \label{L}\ee where $\cal M$ is the energy
scale. Here, the sign before $e^{4\varphi/{\cal M}}\,X$ is
negative. However, as will be showed that this model has not the
ghost and gradient instabilities, since $Q>0$ and $c_s^2>0$. The
evolution of background is determined by \ba &
&\left(-e^{4\varphi/ {\cal M}}+{15\over {\cal M}^8}X^2+{24\over
{\cal M}^7}H{\dot
\varphi}X\right){\ddot \varphi}\nonumber\\
&+& 3\left(-e^{4\varphi/ {\cal
M}}+{3\over {\cal M}^8}X^2\right)H{\dot \varphi}\nonumber\\
&+& \left(-{4\over {\cal M}}e^{4\varphi/ {\cal M}}+ {6{\dot
H}{\dot \varphi}^2\over {\cal M}^7}+{18{ H}^2{\dot \varphi}^2\over
{\cal M}^7}\right)X=0, \label{phi}\ea \be 3H^2 M_P^2 =\,
-\,e^{4\varphi/ {\cal M}}\,X+{5\over {\cal M}^8}X^3 + {6\over
{\cal M}^7}X {\dot \varphi}^3H
.\label{H}\ee  We require that initially $\epsilon\ll -1$, and
behaviors as Eq.(\ref{e2}). This can be found by requiring
$e^{4\varphi/ {\cal M}}X\simeq {5X^3\over {\cal M}^8}$ in
Eq.(\ref{H}).
This gives \be e^{\varphi/{\cal M}}=\left({5\over
4}\right)^{1/4}{1\over {\cal M}(t_*-t)}.\label{ephi}\ee Thus \be
{\dot \varphi}={{\cal M}\over (t_*-t)}.\label{dphi}\ee Thus \be
H\simeq {{\dot \varphi}^5\over {\cal M}^7}\simeq {1\over {\cal
M}^2 M_P^2(t_*-t)^5} \label{H1}\ee 
is induced. Thus for ${\cal M}M_P\sim\Lambda_*^2 $, Eq.(\ref{h})
is obtained. This gives Eq.(\ref{e2}), which is just required
evolution.

Eqs.(\ref{phi}) and (\ref{H}) are numerically solved in Fig.2 and
Fig.3. We can see that Eqs.(\ref{dphi}) and (\ref{H1}) can be
highly consistent with accurate solutions for a long range of
time. The significant deviation only occurs around $t_f\sim {\cal
O}(t_*)$. We might think that the slow expanding phase ends when
the significant deviation appears, and the reheating begins.
However, it might be possible that the reheating of universe
begins some time after the significant deviation occurs, since the
perturbation generated during this period only are the
perturbation on small scale, which has not to be scale invariant.

\begin{figure}[t]
\includegraphics[width=7cm]{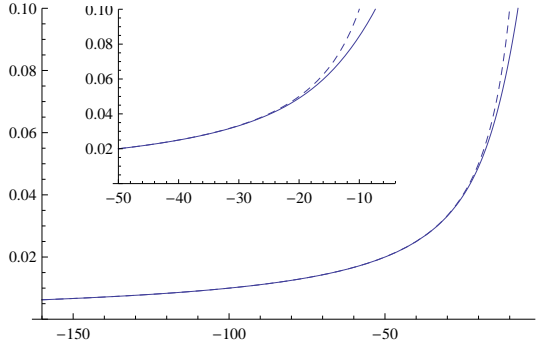}
\caption{ The evolution of $\dot \varphi$ with respect to the
time. The initial values of $\varphi$ and $\dot \varphi$ are
required to satisfy Eqs.(\ref{ephi}) and (\ref{dphi}),
respectively. The parameter ${\cal M}=0.01 M_P$. The dashed line
is that of Eq.(\ref{dphi}). The inset is that around $t_f\sim
{\cal O}(t_*)$. }
\end{figure}

\begin{figure}[t]
\includegraphics[width=7cm]{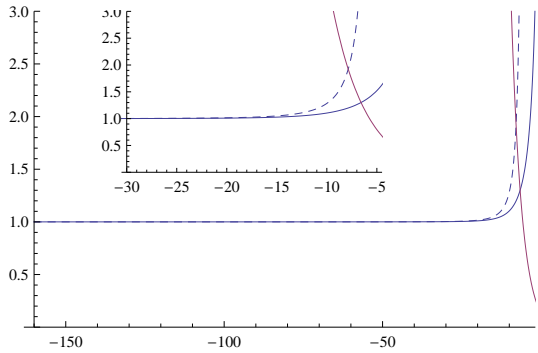}
\caption{ The evolutions of $a$ and $H$ with respect to the time.
The red line is that of $H$. The black line is that of $a$, while
the black dashed line is that of Eq.(\ref{aa}). The inset is that
around $t_f\sim {\cal O}(t_*)$. }
\end{figure}

Eqs.(\ref{dphi}) and (\ref{H1}) implies $H{\dot \varphi}{\cal
M}\ll X$, $H{\dot \varphi}/{\cal M}^3\ll e^{2\varphi/{\cal M}}$,
and $H{\dot \varphi}\ll {\ddot \varphi}$, since \be H\sim {1\over
(t_*-t)^5}\ll {1\over (t_*-t)} \label{HH}\ee for $|\epsilon|\gg
1$, i.e.$\sqrt{{\cal M}M_P}(t_*-t)\gg 1$. Thus Eq.(\ref{phi}) is
approximately \be \left(-e^{4\varphi/ {\cal M}}+{15\over {\cal
M}^8}X^2\right){\ddot \varphi} -{4\over {\cal M}}e^{4\varphi/
{\cal M}}X\simeq 0 \label{phiapp}\ee for $\sqrt{{\cal M}
M_P}(t_*-t)\gg 1$. It can be found that Eq.(\ref{phiapp}) is
consistent with Eqs.(\ref{ephi}) and (\ref{dphi}). Thus the
equation of the perturbation $\delta\varphi$ of $\varphi$ is \ba &
&\,\,\,\,\,\,\,\,\left(-e^{4\varphi/ {\cal M}}+{15\over 4{\cal
M}^8}{\dot \varphi}^4\right)\delta{\ddot \varphi} -{4\over {\cal
M}}e^{4\varphi/ {\cal M}}{\dot\varphi}\delta{\dot
\varphi}\nonumber\\ &+& {15\over {\cal M}^8}{\dot
\varphi}^3{\ddot\varphi}\delta{\dot \varphi} - \left({4\over {\cal
M}}{\ddot \varphi}+{8\over {\cal M}^2}{\dot\varphi}^2
\right)e^{4\varphi/ {\cal M}}\delta\varphi \simeq 0. \nonumber\ea
When Eqs.(\ref{ephi}) and (\ref{dphi}) are considered, the
solution is 
\ba \delta\varphi &\sim &\,\,
(t_*-t)^6, \,\,\, is \,\,\, {\rm decaying}\,\,\, {\rm mode} \\
& or &\,\, { 1/ (t_*-t)}, \,\,\,is \,\,\, {\rm
increasing}\,\,\,{\rm mode}. \label{detal}\ea The decaying mode is
negligible. The increasing mode is dominated. Thus
$\delta\varphi\sim {{\dot \varphi}\over {\cal M}}$. Thus for
${\cal M}\Delta t \gg 1$, $\delta\varphi\ll \Delta\varphi$. Thus
if initially $\delta\varphi\ll \varphi$ is satisfied, it will be
valid all along. When the time arrives around $t_f$, Eq.(\ref{HH})
will be not right. Thus Eq.(\ref{phiapp}) can not be found. This
explains why there will be significant deviation for
Eq.(\ref{dphi}) around $t_f$.

There might be other fluids, However, their energies generally do
not increase, since the expansion is slow. Thus for $|\epsilon|\gg
1$, i.e.$\sqrt{{\cal M}M_P}(t_*-t)\gg 1$, the evolution of
background, given by Eqs.(\ref{dphi}) and (\ref{H1}), is stable.

\subsection{The curvature perturbation}

\begin{figure}[t]
\includegraphics[width=7cm]{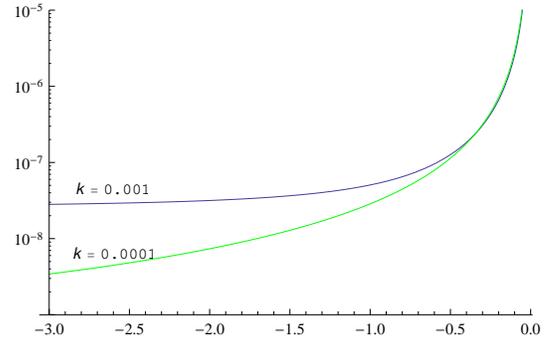}
\caption{ The evolutions of the amplitude of curvature
perturbation for different $k$ with respect to the time. The green
and black lines are that with different $k$. Here, the time axis
is rescale as ${\cal M}t$ for the convenience of numerical
calculation, $t$ is that in Fig.2 and Fig.3, ${\cal M}=0.01$.}
\end{figure}

\begin{figure}[t]
\includegraphics[width=7cm]{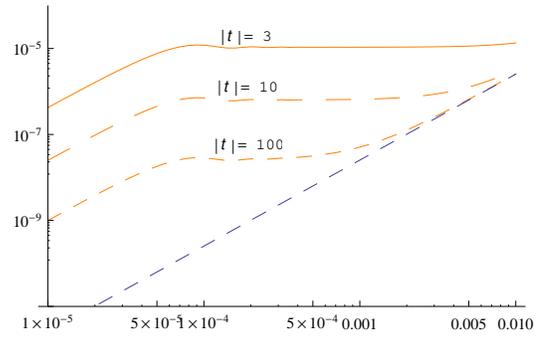}
\caption{ The spectrum of curvature perturbation at different
times with respect to $k$. The black dashed line is initial
spectrum. The short dashed, long dashed and solid orange lines are
the spectra at different times, respectively. There is a cutoff
$k_{cutoff}\sim 5\times 10^{-5}$, below which the spectrum is not
scale invariant, which is explained in the text. }
\end{figure}

$\cal R$ satisfies Eq.(\ref{uk}). We follow the definitions and
calculations of Refs.\cite{KYY},\cite{KYY2}
Here, the generalized Galileon action is (\ref{L}). Thus it is
found that \ba {\cal F}&= &\,-e^{4\varphi/ {\cal M}}+{3X^2\over
{\cal
M}^8}+{8X\over {\cal M}^7}({\ddot \varphi}+H{\dot \varphi})-{8X^4\over {\cal M}^{14}M_P^2} \nonumber\\
&\simeq & {7\over 2{\cal M}^4 (t_*-t)^4}\label{F}\\ {\cal G}&=&
\,-e^{4\varphi/ {\cal M}}+{15X^2\over {\cal M}^8}+{12H{\dot
\varphi}^3\over {\cal M}^7}+{12X^4\over {\cal M}^{14}M_P^2}\nonumber\\
& \simeq & {5\over 2{\cal M}^4 (t_*-t)^4}
 \ea
for ${\cal M}(t_*-t)\gg 1$. In \cite{KYY}, the results are applied
to that of inflation, however, which are actually general for
arbitrary evolution. Thus $Q$ is given by
\be Q = {{\cal F}X \over M_P^2(H-{2{\dot \varphi}X^2\over {\cal
M}^7M_P^2})^2} \sim {M^{14}M_P^2{\cal F}\over {\dot\varphi}^8}
\simeq  {\cal M}^2 M_P^2(t_*-t)^4, \label{Q}\ee where
Eqs.(\ref{dphi}) and (\ref{H1}) are applied. Thus $Q\sim
|\epsilon|
>0$, which is just required here, satisfies Eq.(\ref{e2}). There is not the ghost
instability. Here, the importance of $X^2\Box\varphi$ is obvious,
because if it disappears in (\ref{L}), $\cal F$ is given by \be
{\cal F}=\,-e^{4\varphi/ {\cal M}}+{3X^2\over {\cal M}^8}\simeq
-{1\over 2{\cal M}^4 (t_*-t)^4}<0, \ee $Q>0$ will hardly be
obtained, which is consistent with $Q=\epsilon<0$ for this case.
This indicate that it is $X^2\Box\varphi$ that alters the sign of
$Q$, and leads $Q\sim |\epsilon|
>0$. The $c_s^2$ is given by \be c_s^2={{\cal F}\over
{\cal G}}\sim 1.4. \label{cs}\ee Thus $c_s^2>0 $ is constant,
which is also just required. The sign of $c_s^2$ is determined by
the signs of $\cal F$ and $\cal G$, both are positive. Here,
obviously ${\cal F}>0$ is also required to assure $c_s^2>0$.
Thus there are not the ghost and gradient instabilities, the
effective theory is healthy.

We plot the evolution of the amplitude of the curvature
perturbation in Fig.4, and the spectrum of perturbation in Fig.5.
We can see that the perturbation is initially not increasing,
since it is inside the $\cal R$ horizon. The increase begins until
the perturbation mode leaves the $\cal R$ horizon. The longer the
wavelength of perturbation is, the earlier the perturbation leaves
the $\cal R$ horizon, the earlier it begins to increase. However,
since the shorter the wavelength of perturbation is, the larger
its initial amplitude is, all perturbation modes will eventually
have same amplitude.

There is a cutoff $k_{cutoff}$ in Fig.5, which is given by \be
k_{cutoff}\sim {\cal H}_{inifr}, \ee where ${\cal H}_{inifr}$ is
${\cal H}_{freeze}$ at initial time, and can be changed with the
difference of the initial parameters in the numerical calculation.
The spectrum is scale invariant for $k>k_{cutoff}$. However, for
$k<k_{cutoff}$, since the corresponding perturbation modes are
outside the $\cal R$ horizon all along, only are their amplitudes
increasing but not the shape of the spectrum is not altered
\cite{Piao0901},\cite{ZLP}.

The spectrum of $\cal R$ is scale invariant. The amplitude of
spectrum is given by Eq.(\ref{P1}) \be {\cal P}_{\cal R}^{1/2}\sim
\sqrt{{\cal M}\over c_s M_P}, \ee where $\Lambda_*\sim\sqrt{{\cal
M}M_P}$ is applied. Thus ${\cal P}_{\cal R}^{1/2}\sim 10^{-5}$
requires ${\cal M}\sim 10^{-10} c_sM_P$. Thus ${\cal M}\sim
10^9$Gev for $c_s\simeq 1$. The only adjusted parameter in this
model is fixed by the observation. There is not other finetuning.





\subsection{The reheating}

When the slowly expanding phase ends, the energy of Galileon field
is required to be released into the radiation, and the universe
reheats. Hereafter, the evolution of hot ``big bang" cosmology
begins. We can notice that before this, the perturbation mode has
leaved the Hubble horizon.

Here, in certain sense, the reheating is alike with that for
inflation. The preheating theory after inflation has been
developed in \cite{KLS2},\cite{TB}. In general, during the
preheating phase after inflation the energy of inflaton will be
rapidly released
by the parametric resonance effects, due to the coupling of
inflaton with other fields. Then this issue has been extensively
studied, see \cite{BTW},\cite{ABCM},\cite{MR} for reviews.

We will apply the instant preheating mechanism \cite{FKL} for
given case here. We consider the straight coupling of $\varphi$
with $\chi$ particle as \be {\cal L}\sim
g^2(\varphi-\varphi_{reh})^2\chi^2, \label{R1}\ee where $g$ is the
coupling constant. The effective mass of $\chi$ particle is
$M_{\chi eff}^2\sim g^2(\varphi-\varphi_{reh})^2$. When the
$\varphi$ field arrives at the region around $ \varphi_{reh}$,
$M_{\chi eff}^2\lesssim {\dot M}_{\chi eff}$, the adiabatic
condition is broke, and the productions of $\chi$ particles will
be inevitable. This generally occurs in a region around
$\varphi_{reh}$, $\Delta\varphi\lesssim{\dot \varphi}_{reh}/g$, in
which ${\dot \varphi}_{reh}$ is the velocity of $\varphi$ through
$\varphi_{reh}$. Thus the productions of $\chi$ particles is
instantaneous, $\Delta t_{reh}\sim 1/\sqrt{g{\dot
\varphi}_{reh}}$.

\begin{figure}[t]
\begin{center}
\includegraphics[width=7cm]{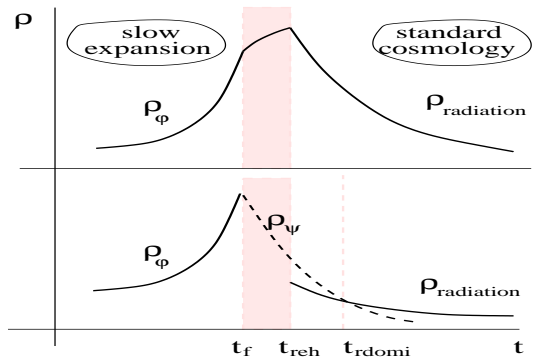}
\caption{ The sketch of the evolution of the energy density $\rho$
for different reheating courses discussed here. } \label{fig1}
\end{center}
\end{figure}

The number density $n_{\chi}$ of $\chi$ particle is \be
n_{\chi}={1\over 2\pi^2}\int n_k k^2 dk\simeq {g^{3/2}{\dot
\varphi}_{reh}^{3/2}\over 8\pi^3}, \ee where $n_k$ is the
occupation number of $\chi$ particle. Thus
$\rho_{\chi}=n_{\chi}M_{\chi} \sim g^{2}{\dot \varphi}_{reh}^{2}$,
since $M_{\chi eff}\sim g(\varphi-\varphi_{reh})\sim g{\dot
\varphi}_{reh}\Delta t_{reh}$. Thus the energy drained by the
production of $\chi$ particle is \be {\rho_{\chi}\over
\rho_{\varphi reh}}
\sim {g^2\over 8\pi^3}{\cal M}^6 M_P^2(t_*-t_{reh})^8, \ee where
Eqs.(\ref{dphi}) and (\ref{H1}) are applied, and $\rho_{\varphi
reh}$ is the energy density of $\varphi$ around $t_{reh}$. We
assume $t_f\sim t_{reh}$ for simplicity, i.e. the reheating occurs
at the time when the slow expansion ends. Thus ${\cal M}^2
M_P^2(t_*-t_{reh})^4\sim 1$. This implies \be {\rho_{\chi}\over
\rho_{\varphi reh}}\sim {g^2{\cal M}^2\over 8\pi^3 M_P^2}.
\label{rhorho}\ee We generally require ${\cal M}\ll 1$ and $g<1$.
Thus ${\rho_{\chi}/ \rho_{\varphi reh}} \ll 1$, which indicates
that for such a single preheating, the energy of $\varphi$ can
hardly be released completely, the universe is still dominated by
$\rho_{\varphi}$, which will continue all along, since the energy
density of $\varphi$ is increasing with the expansion of universe
while that of $\chi$ particle is decreasing.

However, there might be $\cal N$ couplings, one of which is alike
with (\ref{R1}). We can find, after doing similar calculations,
that when \be {\cal N}> {M_P^2\over g^2{\cal M}^2}, \ee the
release of the energy of $\varphi$ will be complete. The sketch of
this reheating course is plotted in upper panel in Fig.6. We
assume that the $\chi$ particle produced is rapidly transferred
into the radiation. In this case, the reheating temperature $T_r$
is approximately determined by $\rho_{\varphi reh}\sim T_{r}^4$.
Thus we have \be T_{r}\sim \left({{\dot \varphi}^{10}\over {\cal
M}^{14} M_P^2}\right)^{1/4}\sim {\cal M}^{1/4}M_P^{3/4},
\label{T}\ee where
${\cal M}^2 M_P^2 (t_*-t_{reh})^4\sim 1$ is applied again. 
Thus if ${\cal M}\sim 10^{-10}M_P$, we have $T_{r}\sim
10^{15}$Gev.

Here, ${\cal N}\gg 1$ is feasible, however, might be
uncomfortable. ${\cal N}\gg 1$ is required is because the energy
of $\varphi$ has to be released completely for one time, or since
the energy density of $\varphi$ is increasing, the universe will
dominated by $\varphi$ all along. However, we also could consider
another channel of the reheating, likes that in phantom inflation.
The energy of $\varphi$ is firstly shifted to the kinetic energy
of a normal field, e.g.$\psi$, and then the energy of $\psi$ is
released by the instant preheating. The sketch of this reheating
course is plotted in lower panel in Fig.6. Here, the energy of
$\psi$ is not required to be released completely, since
$\rho_{\psi}\sim 1/a^6$ is decreasing faster than that of the
radiation, the universe will be dominated by that of the radiation
early or late.

We can implement it by considering a potential of $\varphi$,
illustrated in Fig.7. We require that it is only significant
around or after $|\epsilon|\sim 1$, and is negligible
$|\epsilon|\gg 1$. Then we introduce a waterfall field $\psi$,
coupled to $\varphi$. The effective mass of $\psi$ is initially
positive and becomes negative around $|\epsilon|\sim 1$. Thus
$\psi$ will roll down along its potential. Thus almost all energy
of $\varphi$ will be shifted to $\rho_{\psi}\sim {\dot \psi}^2$.
This energy will be expected to be released by the instant
reheating. Thus there could be a suitable reheating after the slow
expansion ends, after which the evolution of hot ``big bang"
cosmology begins.

\begin{figure}[t]
\begin{center}
\includegraphics[width=7cm]{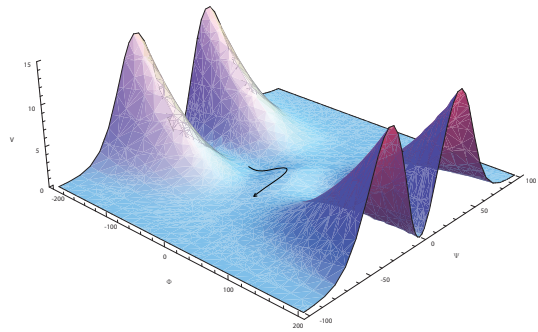}
\caption{ The figure of the effective potential for the exiting
from the slow expansion. The black solid line is the motive
trajectory of field in $(\varphi,\psi)$ space. } \label{fig1}
\end{center}
\end{figure}

\section{Discussion}

When initially $\epsilon\ll -1$ and is rapidly increasing, the
universe is slowly expanding. The spectrum of curvature
perturbation generated during such a phase of slow expansion can
be scale invariant. This provides a mechanism by which an
alternative scenario of early universe can be imagined. Here, we
show a model of such a scenario by applying an effective action of
generalized Galileon.

In principle, $\epsilon<0$ implies the ghost instability. However,
in this model, because of the introduction of Galileon field,
there is not the ghost instability, the perturbation theory is
healthy. In Refs.\cite{PZhou},\cite{Piao1012}, the phantom was
applied for an implementing of slow expansion. In the calculations
of perturbation, for consistence, $|\epsilon|$ is used, though the
initial value of perturbation is still pathologically defined.
However, in the model given here, it can be found that actually
$Q\simeq |\epsilon|$. This in certain sense validates the argument
and calculations used in \cite{PZhou},\cite{Piao1012}, i.e. the
phantom field might be a simple simulation of a full theory
without the ghost below certain physical cutoff, which can give
same results with that of a full theory, when the replacement of
$\epsilon$ with $|\epsilon|$ is done.

When $\epsilon\sim -1$, the slow expansion ends. The exiting to a
hot universe is only a simple reheating, since the universe
expands all along. Thus there is not the problem how the bouncing
is implemented in bouncing cosmologies
\cite{KOS},\cite{GV},\cite{Cai0704}. We have discussed possible
implements of reheating, and found that the available energy of
Galileon field can completely released, the universe can reheat to
a suitable temperature. Thus the model of the slow expansion given
here might be a viable design of the early universe.

The material compares of model with the observations is certainly
interesting, which will place rigid constrains for the model. The
results obtained will be expected to either improves or rules out
this model. We will investigate it elsewhere. However, it should
be pointed that we only bring one of all possible implements of
the slow expansion. In principle, there might be other effective
actions of generalized Galileon, or modified gravity, which could
give the same evolution of background. Thus for the slow
expansion, it might be also significant to find alternative
implements to the model given here, which will help to uplift the
flexility of the slow expansion to the observations.

Here, the scale factor is asymptotic to a constant value in
infinite past, there is not singularity point. Thus in certain
sense, the slow expansion scenario brings a solution to the
cosmological singularity problem. However, it also can be imagined
that after the available energy of the field is released, it might
be placed again in the bottom of its effective potential, and
after the universe undergoes the radiation and matter periods, the
field might dominate again and roll again with increasing energy.
This models an eternally expanding cyclic universe
\cite{Feng06},\cite{Xiong08},\cite{IBF}, i.e. $H$ oscillates
periodically while $a$ expands all along. The implement of this
cyclic universe might be interesting for refining with the model
given here.

Here, $c_s$ is constant is set. However, its change will obviously
enlarge the space of solutions of the scale invariance of
curvature perturbation
\cite{Picon},\cite{Piao0609},\cite{JM},\cite{KP},\cite{Kinney}. In
certain sense all possibilities of the changes of $a$, $Q$ and
$c_s^2$ might
be interesting for further exploring.



\textbf{Acknowledgments} This work is supported in part by NSFC
under Grant No:10775180, 11075205, in part by the Scientific
Research Fund of GUCAS(NO:055101BM03), in part by National Basic
Research Program of China, No:2010CB832804.

\end{document}